\documentclass[12pt]{article}
\usepackage{ctex}
\usepackage{amssymb,amsmath,color,mathdots}
\usepackage[colorlinks,linkcolor=blue]{hyperref}
\usepackage{geometry}
\usepackage{tabularx}
\usepackage{bm}
\usepackage{authblk}
\usepackage[square, comma, sort&compress, numbers]{natbib}

	\title{\bf The $b$-weight distribution for MDS codes}	
		\author{\small Canze Zhu}
	\author{\small Qunying Liao
		\thanks{Corresponding author.\\
			{E-mail. qunyingliao@sicnu.edu.cn (Q. Liao), ~canzezhu@163.com (C. Zhu).}	\\				
			{~Supported by National Natural Science Foundation of China (Grant No. 12071321).}}
		}
	\affil[]{\small (College of Mathematical Science, Sichuan Normal University, Chengdu Sichuan, 610066)}
	\date{}
	\newtheorem{theorem}{Theorem}[section]
	
	\newtheorem{lemma}{Lemma}[section]
	\newtheorem{example}{Example}[section]
	
	\newtheorem{corollary}{Corollary}[section]
	\newtheorem{remark}{Remark}[section]

	\catcode`@=11 \@addtoreset{equation}{section} \catcode`@=12
\begin{document}
	\maketitle
	{\bf Abstract.}
	{\small For a positive integer $b\ge2$, the $b$-symbol code is a new coding framework proposed to combat $b$-errors in $b$-symbol
		 read channels. Especially, the $2$-symbol code is called a symbol-pair code. Remarkably,  a classical maximum distance separable (MDS) code is
		also an MDS $b$-symbol code. Recently, for any MDS code $\mathcal{C}$, Ma and Luo determined the symbol-pair weight distribution of $\mathcal{C}$. In this paper, by calculating the number of solutions for some equations and utilizing some shortened codes of $\mathcal{C}$, we give the connection between the $b$-weight distribution and the number of codewords in shortened codes of $\mathcal{C}$ with special shape. Furthermore, note that shortened codes of $\mathcal{C}$ are also MDS codes,  the number of these codewords with special shape  are also determined by the shorten method. From the above calculation, the $b$-weight distribution of $\mathcal{C}$ is determined. Our result generalies the corresonding result of Ma and Luo \cite{ML}.\\

	

	\section{Introduction}
	In 2011, Cassuto and Blaum \cite{C2} first proposed a new coding framework for symbol-pair read channels. The outputs of the read process in the channels are overlapping pairs of
		symbols. After that, Chee et al. \cite{C4,C5} established a Singleton-type bound for symbol-pair codes and considered the constructions for symbol-pair codes meeting the bound.
		 
	In 2016, Yaakobi et al. \cite{Y23} generalized the coding framework for symbol-pair read channels to that for $b$-symbol read channels, where the read operation is performed as a
		consecutive sequence of $b$-symbols $(b>2)$. They also generalized some known results for symbol-pair read channels to those for $b$-symbol read channels, and showed that a
		code $\mathcal{C}$ with minimum $b$-distance $d_b$ can correct at most $\lfloor\frac{d_b-2}{2}\rfloor$ $b$-symbol errors. Furthermore, they gave a decoding algorithm based on a bounded distance decoder for the cyclic code. Later, Ding et al. \cite{D8} extended the Singleton-type bound to be the $b$-symbol case and constructed MDS $b$-symbol codes basing on projective geometry, and then, they showed that an MDS $b$-symbol code with $b$-distance $d_b<n$ is also an MDS $(b+1)$-symbol code.  Moreover, relative results about constructions for MDS symbol-pair codes can be seen in \cite{C6,D7,D9,D12,E14,K17,K18,L19}.
	
	It is well-known that the Hamming weight distribution for an MDS code can be uniquely determined \cite{H16}. However, there are few works on the $b$-weight distribution for  MDS $b$-symbol codes. In 2021, Ma and Luo  gave the symbol-pair weight distribution of MDS codes by employing some shortened MDS codes \cite{ML}. Furthermore, there are some results about the $b$-weight distribution for cyclic code. In 2018, Sun et al. gave the symbol-pair weight distribution of a class of repeated-root cyclic codes \cite{S1}. In 2021, Shi et al. presented the geometric approach to $b$-weights of cyclic codes, and  constructed a class of irreducible cyclic codes with constant $b$-weight \cite{S2}. Recently, Zhu et al. gave a complete $b$-weight distribution of a class of irreducible cyclic codes with two nonzero $b$-weight \cite{Z}.
	 
	In this paper, for any MDS code $\mathcal{C}$,  the connection between the $b$-weight distribution and the number of codewords in shortened codes of $\mathcal{C}$ with special shape is presented by calculating the number of solutions for some equations and utilizing some shortened codes of $\mathcal{C}$. Furthermore, by the shorten method, the number of these codewords with special shape is given. And then the $b$-weight distribution is completely determined.   Especially, for $b=2$, we obtain Theorem $1$ in \cite{ML}.
	
	The paper is organized as follows. In section 2, some related basic notations and
	results are given. In section 3, the main results are presented. In section 4, the proofs
	for the main results are given. In section 5, we conclude the whole paper.
	\section{Preliminaries}
	Throughout this paper, let $\mathbb{F}_q$ be the finite field with $q$ elements, where $q$ is a prime power. An element in $\mathbb{F}_q$ is also called a symbol. Let $b$ be a positive integer. For a vector $\mathbf{x} = (x_1 ,x_2,\ldots, x_{n})\in\mathbb{F}_q^n$, we define the $b$-symbol read vector
	of $\mathbf{x}$ as
	\begin{align*}
		\pi_b (\mathbf{x}) = ((x_1,x_2,\ldots,x_{b}), (x_2,x_3,\ldots,x_{b+1}),\ldots ,(x_{n-1}, x_n,\ldots,x_{b-2}),(x_{n}, x_1,\ldots,x_{b-1})).
	\end{align*}
	For any two vectors $\mathbf{x}$ and $\mathbf{y}$ in $\mathbb{F}_q^n$, one has
		\begin{align*}
	\pi_b (\mathbf{x}+\mathbf{y}) =\pi_b ( \mathbf{x})+\pi_b ( \mathbf{y}),
	\end{align*}
	and the $b$-distance between $\mathbf{x}$ and $\mathbf{y}$ is defined as
	\begin{align*}
		D_b(\mathbf{x},\mathbf{y}):=\#\{1\le i\le n~|~(x_i,\ldots,x_{i+b-1})\neq (y_i,\ldots,y_{i+b-1}) \},
	\end{align*}
	where $x_m=x_{m-n}$ if $m>n$. Accordingly, the $b$-weight of $\mathbf{x}$ is defined as
	\begin{align*}
		wt_b(\mathbf{x}):=\#\{1\le i\le n~|~(x_i,\ldots,x_{i+b-1})\neq\mathbf{0} \},
	\end{align*}
	where $\mathbf{0}$  denotes the all-zeros vector in $\mathbb{F}_q^n$,  and $x_m=x_{m-n}$ if $m>n$. Especially, for $b=1$, $D_b(\mathbf{x},\mathbf{y})$ and $wt_b(\mathbf{x})$ are called Hamming distance and Hamming weight, respectively.
	
	An  linear $b$-symbol code $\mathcal{C}$ over $\mathbb{F}_q$ is a $k$-dimensional subspace of $\mathbb{F}_q^n$ with minimum $b$-distance
	 \begin{align*}
	 d_b:=\min\#\{D_b(\mathbf{x},\mathbf{y})~|~\mathbf{x},~\mathbf{y} \in \mathcal{C},~  \mathbf{x}\neq\mathbf{y}  \}.
	 \end{align*}
	 Especially, if $b=1$, let $d=d_1$, then $\mathcal{C}$ is called an $[n,k,d]$ linear code.\\
	 
	The size of a $b$-symbol code satisfies the following Singleton bound.
	 \begin{lemma}\cite{D8}
	 	 Let $b \le d_b \le n$. If $\mathcal{C}$ is a $b$-symbol code over $\mathbb{F}_q$ with length $n$ and
	 	minimum $b$-distance $d_b$, then 
	 	\begin{align*}
	 		\big| \mathcal{C} \big| \le q^{n-d_b +b}.
	 	\end{align*}
	 \end{lemma}
 
 	An $b$-symbol code with parameters achieving the Singleton bound is called a maximum distance separable (in short, MDS) $b$-symbol code. For $b=1$, the MDS $b$-symbol code is also called the MDS code. It is easy to see that an MDS code is also an MDS $b$-symbol code by the following lemma.
 	\begin{lemma}\cite{D8}
 		An MDS $b$-symbol code with $d_b<n$ is an MDS $(b+1)$-symbol code.
 	\end{lemma}
 
 Next, we give some notations on shortened codes and the Hamming weight distribution for MDS codes.
 	
 	Suppose $\mathcal{C}$ is a code of length $n$ over $\mathbb{F}_q$, we denote $[1,n]=\{1,2,\ldots,n\}$, and for $S\subseteq[1,n]$, denote $\overline {S}=[1,n]\backslash S$. Then the code $\mathcal{C}_S$ shortened on $S$ from $\mathcal{C}$ is defined by
 	\begin{align*}
 	\mathcal{C}_S=\bigg\{(c_j)_{j\in\overline{S}} ~\bigg|~\text{there exists some ~}(c_1,c_1,\ldots,c_{n})\in\mathcal{C}, \text{such that}~c_i=0~\text{for all}~i\in S\bigg\}.
 	\end{align*}For more details, see Section 1.5 in \cite{H16}.
 	
 	The Hamming weight distribution for the MDS code is determined by the following lemma.
 	\begin{lemma}\cite{H16}\label{MDS}
 		Let $\mathcal{C}$ be an $[n, k, d]$ MDS code and $A_i$ $(i=0,1,\ldots,n)$ the number of codewords in $\mathcal{C}$
 		with weight $i$. Then the weight distribution for $\mathcal{C}$ is given by $A_0 = 1$, $A_i = 0~(1 \le i \le d-1)$
 		and
 		\begin{align*}
 		A_i=\binom{n}{i}\sum\limits_{j=0}^{i-d}(-1)^j\binom{i}{j}\big(q^{i+1-d-j}-1\big)\quad(d\le i\le n).
 		\end{align*}
 	\end{lemma}
 
  The following lemma characterizes that a code shortened on $S$ from an MDS code is also an MDS code.
 	\begin{lemma}\cite{H16}\label{MDSS}
 		Let $\mathcal{C}$ be an $[n, k, n-k + 1]_q$ MDS code  and $\mathcal{C}_S$ be the code shortened on $S$ from $\mathcal{C}$, where $S$ is a set with $|S|=s<k$. Then $\mathcal{C}_S$  is an $[n-s, k-s, n-k+1]_q$ MDS code.
 	\end{lemma}
 
 The following notations $N_\infty(r,L)$ and $N_b(r,L)$ are useful to calculate the number of codewords in an MDS code with special shape. For $b$, $r$, $L$ $\in\mathbb{N}^{*}$ with $b\ge3$ and $r\le L$, let $N_\infty(r,L)$ be the number of solutions for the equation
 \begin{align*}
 \begin{cases}
 x_1+x_2+\cdots+x_r=L;\\
 x_i\in\mathbb{N}^{*}.
 \end{cases}
 \end{align*}
 Let $N_b(e,L)$ be the number of solutions for equation
 \begin{align*}
 	\begin{cases}
 	x_1+x_2+\cdots+x_r=L;\\
 	x_i\in\mathbb{N}^{*},~1\le x_i\le b-2~(i=1,\ldots,r),
 	\end{cases}
 \end{align*}
furthermore, we set $N_{b}(0,0)=1$ for convience.  In fact, $N_\infty(e,L)$ and $N_b(e,L)$ are correspond to the number of $r$-combinations of a multiset problem.  $N_\infty(e,L)$ is given in the following 
	\begin{lemma}[Chapter $6.2$ \cite{B1977}]\label{nc}
	$N_\infty(e,L)=\binom{L-1}{e-1}$.
	\end{lemma}
In addition, $N_b(e,L)$ can be calculated by $N_\infty(e,L)$ and the inclusion-exclusion principle, for more details, see Chapter $2.5$ and  Chapter $6.2$ \cite{B1977}. 

	\section{The $b$-weight distribution for MDS codes}
	
	In this section, the  $b$-weight distribution for MDS codes is determined. We begin with the following notation.
	
	{\bf  Notation:} For any positive integers $d,I,L$ and $L_i~(i=1,\ldots,I)$ with $\sum\limits_{i=1}^{I}L_i=L>d$,  let $\mathcal{C}$ be an  MDS code $[L,L-d+1,d]$. Let $F_{b,d}(L_1,\ldots,L_I)$ be the number of codewords $\mathbf{c}\in\mathcal{C}$ satisfying the following two conditions.
	
	$(\mathrm{F}1)$ $\mathbf{c} =(\underbrace{c_1,\ldots,c_{L_1}}_{1},\underbrace{c_{L_1+1},\ldots,c_{L_1+L_2}}_{2},\ldots,\underbrace{c_{L_1+\cdots+L_{I-1}+1},\ldots,c_{L_1+\cdots+L_I}}_{I})$ with \\
	\begin{align*}
	c_{L_1+\cdots+L_{i-1}+1}\neq 0\quad\text{and}\quad c_{{L_1+\cdots+L_i}}\neq 0~~(1 \le i\le I);
	\end{align*}
	
	$(\mathrm{F}2)$ $wt_{b-1}(\mathbf{c})=L$.
	
	\noindent For completeness, if $I=1$, we set $L_1+\cdots+L_{I-1}=0$ in $(\mathrm{F}1)$.\\
	
	In the following theorems, we set $\binom{m_1}{0}=1$ $(m_1\in\mathbb{N})$ and $\binom{0}{m_2}=0$ $(m_2\in\mathbb{N}^{*})$ for completeness.	 Now the connnection between the $b$-weight distribution for an MDS code and $F_{b,d}(L_1,\ldots,L_I)$ is presented.

	\begin{theorem}\label{t31}
		 For $b\ge 2$ and an $[n,k,d]$ MDS code $\mathcal{C}$, let $A^b(w)$ be the number of codewords in $\mathcal{C}$ with $b$-weight $w$. If $d+b-1<n$, then
		\begin{align*}
			A^b(w)=\begin{cases}
			1,\quad&w=0;\\
			0,\quad& 1\le w\le d+b-2;
			\end{cases}
		\end{align*}
		and for $d+b-1\le w\le n$,
		{\small \begin{align*}
			A^b(w)=&\delta_n(w) \sum\limits_{t=0}^{b-2}(t+1)F_{b,d}(n-t)+(n-w+b)F_{b,d}(w-b+1)\\
			&+	\sum\limits_{t=0}^{b-2}(t+1)\sum\limits_{i=1}^{\lfloor M_1(w,b,t)\rfloor}\binom{n-w+i-1}{i-1}\sum\limits_{\substack{a_1,\ldots,a_{i+1}\in \mathbb{N}^{*}\\a_1+\cdots+a_{i+1}=w-t-i(b-1)}}F_{b,d}(a_1,\ldots,a_{i+1})\\
			&+\sum\limits_{t=b-1}^{n-w+b-1}(t+1)\sum\limits_{i=1}^{\lfloor M_2(w,b)\rfloor}\binom{n-w-t+b+i-2}{i-1}\!\!\!\!\!\!\!\!\!\!\!\!\sum\limits_{ \substack{a_1,\ldots,a_{i+1}\in \mathbb{N}^{*}\\a_1+\cdots+a_{i+1}=w-(i+1)(b-1)}}\!\!\!\!\!\!\!\!\!\!\!\!F_{b,d}(a_1,\ldots,a_{i+1}),
			\end{align*}}
		where~
		
	\noindent	$~~~\delta_n(w)=\begin{cases}
			1,\!&w=n;\\
			0,\!&\text{otherwise},
			\end{cases}$~~$M_1(w,b,t)=\min \big\{\frac{w-t-1}{b},\frac{w-t-d}{b-1}\big\},$~~$M_2(w,b)=\min \big\{ \frac{w-b}{b},\frac{w-d}{b-1}-1\big\}.$
\end{theorem}

In the following theorem, we show that the value of $F_{b,d}(L_1,\ldots,L_I)$ can be determined by $N_{b}(r,L)$.
	\begin{theorem}\label{t32}
		\begin{align*} F_{b,d}(L_1,\ldots,L_I)
		 =\begin{cases}
		 	A\Big(\sum\limits_{i=1}^{I}L_i,d\Big),\quad&\text{if}~b=2;\\
		 f_b(L_1,m_1)\cdots f_b(L_I,m_I)A\Big(\sum\limits_{i=1}^{I}m_i,d\Big),\quad&\text{if}~b\ge 3.\\
		 \end{cases} 
	 \end{align*}
 where
\begin{align*}
f_b(L_i,m_i)= \sum\limits_{m_i=\big\lceil \frac{L_i+(b-2)}{b-1}\big\rceil}^{L_i}\sum\limits_{e_i=\big\lceil\frac{L_i-m_i}{b-2}\big\rceil}^{\min\{m_i-1,L_i-m_i\}}\binom{m_i-1}{e_i}N_{b}(e_i,L_i-m_i), 
\end{align*}
and
{\begin{align*}
	\begin{aligned}
	A\Big(\sum\limits_{i=1}^{I}m_i,d\Big)=\begin{cases}
	0, &\sum\limits_{i=1}^{I}m_i<d;\\
	\sum\limits_{j=0}^{\sum\limits_{i=1}^{I}m_i-d}(-1)^j\binom{\sum\limits_{i=1}^{I}m_i}{j}\bigg(q^{\sum\limits_{i=1}^{I}m_i+1-d-j}-1\bigg), &\text{otherwise}.
	\end{cases}
	\end{aligned}
\end{align*}}
\end{theorem}

	
\begin{remark}
	 By  Theorems \ref{t31}-\ref{t32}, the $b$-weight distribution for an MDS code can be determined.
 \end{remark}

\begin{remark}
	In general, the expression of $A^{b}(w)$ $(d+b-1\le w\le n)$ is long and complex, the main reason is that the value of the term $\sum\limits_{\substack{a_1,\ldots,a_{i+1}\in \mathbb{N}^{*}\\a_1+\cdots+a_{i+1}=w-t-i(b-1)}}F_{b,d}(a_1,\ldots,a_{i+1})$ is associated with $N_b(e,L)$, the value of which is not explicted. However, not only for $b=2~\text{or}~3$, but also for some special cases of general $b$, $N_b(e,L)$ can be given explictly, thus $A^{b}(w)$ $(d+b-1\le w\le n)$ can be given conciesly, see corollaries \ref{c21}-\ref{C4}.
\end{remark}
	\begin{corollary}[Theorem $1$, \cite{ML}]\label{c21}
		Let $\mathcal{C}$ be an $[n,k,d]$ MDS code,  then
			\begin{align*}
		A^2(w)=\begin{cases}
		1,\quad &w=0;\\
		0,\quad &1\le w\le d;
		\end{cases}
		\end{align*}
		and for $d+1\le w\le n$,
		\begin{align*}
		A^2(w)
		=&\delta_n(w)A(n,d)
		+\sum\limits_{i=1}^{\lfloor M_1(w,0,2)\rfloor}\binom{n-w+i-1}{i-1}\binom{w-i-1}{i}A(w-i,d)\\
		&+\sum\limits_{i=1}^{\lfloor M_2(w,2)\rfloor+1}\bigg(2\binom{n-w+i-1}{i-1}+\binom{n-w+i}{i}\bigg)\binom{w-i-1}{i-1}A(w-i,d).
		\end{align*}\\ 
	\end{corollary}

{\bf Proof}. By $$\sum\limits_{ \substack{a_1,\ldots,a_{i+1}\in \mathbb{N}^{*}\\a_1+\cdots+a_{i+1}=N}}1=\binom{N-1}{i} ~~\big(N=w-i~\text{or}~w-(i+1)\big)$$
and 
$$\sum\limits_{t=1}^{n-w+1}(t+1)\binom{n-w-t+i}{i-1}=2\binom{n-w+i}{i}+\binom{n-w+i}{i+1},$$
let $b=2$ in Theorems \ref{t31}-\ref{t32}, then we get the corollary.$\hfill\Box$\\

	\begin{corollary}
			 For an $[n,k,d]$ MDS code $\mathcal{C}$ with $d+2=n$, we have $A^3(w)=\begin{cases}
		1,\quad&w=0;\\
		0,\quad& 1\le w\le d+1,
	 \end{cases}$
	 
\noindent		and for $d+2\le w\le n$,
		{\small \begin{align*}
			A^3(w)=&\delta_n(w) \sum\limits_{t=0}^{1}(t+1)F_{b,d}(n-t)+(n+3-w)F_{3,d}(w-2)\\
			&+	\sum\limits_{t=0}^{1}(t+1)\sum\limits_{i=1}^{\lfloor M_1(w,3,t)\rfloor}\binom{n-w+i-1}{i-1}\sum\limits_{\substack{a_1,\ldots,a_{i+1}\in \mathbb{N}^{*}\\a_1+\cdots+a_{i+1}=w-t-2i}}F_{3,d}(a_1,\ldots,a_{i+1})\\
			&+\sum\limits_{t=2}^{n-w+2}(t+1)\sum\limits_{i=1}^{\lfloor M_2(w,3)\rfloor}\binom{n-w-t+i-1}{i-1}\!\!\!\!\!\!\!\!\!\!\!\!\sum\limits_{ \substack{a_1,\ldots,a_{i+1}\in \mathbb{N}^{*}\\a_1+\cdots+a_{i+1}=w-2(i+1)}}\!\!\!\!\!\!\!\!\!\!\!\!F_{3,d}(a_1,\ldots,a_{i+1}),
			\end{align*}}
		where
		\begin{align*}
		 F_{b,d}(L_1,\ldots,L_I)=\sum\limits_{m_1=\big\lceil \frac{L_1+1}{2}\big\rceil}^{L_1}\binom{m_1-1}{L_1-m_1}\cdots \sum\limits_{m_I=\big\lceil \frac{L_{I}+1}{2}\big\rceil}^{L_{I}}\binom{m_I-1}{L_I-m_I} 	A\Big(\sum\limits_{i=1}^{I}m_i,d\Big)
    	\end{align*}
	\end{corollary}

	{\bf Proof}.  Let $b=3$ in Theorems \ref{t31}-\ref{t32}, by the definition of $N_{3}(e_i,L_i-m_i)$, we have
	 $$N_{3}(e_i,L_i-m_i)=\begin{cases}
	1,&\text{~if~} e_i=L_i-m_i;\\
	0,&\text{~if~} e_i\neq L_i-m_i,
	\end{cases}$$  
    and then the $3$-weight distribution for the MDS code is given. $\hfill\Box$\\

\begin{corollary}\label{C3}
	Let $\mathcal{C}$ be an $[n,k,d]$ MDS code, if $w=d+b-1<n$, then $$A^{b}(w)=n(q-1).$$
\end{corollary}

{\bf Proof}. For $w=d+b-1<n$, note that 
\begin{align*}
\delta_n(w)=0,~~\lfloor M_1(w,b,0)\rfloor=1,~~\lfloor M_1(w,b,t)\rfloor\le0~(t\ge 1),~~M_2(w,b)=0，
\end{align*}
By Theorem \ref{t31}, we have
\begin{align*}
A^{b}(d+b-1)=(n-d+1)F_{b,d}(d)+\sum\limits_{\substack{a_1,a_{2}\in \mathbb{N}^{*}\\a_1+a_{2}=d}}F_{b,d}(a_1,a_{2})
\end{align*}
Now by Theorem \ref{t32}, one has
\begin{align*}
~~~~F_{b,d}(d)=A(d,d)=q-1\text{~~~and~~~}F_{b,d}(a_1,a_{2})=A(d,d)=q-1~(a_1+a_2=d).
\end{align*}
Thus
\begin{align*}
A^{b}(d+b-1)=(n-d+1)(q-1)+(d-1)(q-1)=n(q-1).
\end{align*}
$\hfill\Box$\\

\begin{corollary}\label{C4}
For $b\ge 3$ and $d\ge 3$, let $\mathcal{C}$ be an $[d+b,b+1,d]$ MDS code,  then \begin{align}\label{C5}
A^{b}(d+b)=
\sum\limits_{t=0}^{b-1}(t+1)F_{b,d}(n-t)+dA(d+1,d)+(d-1)dA(d,d).
\end{align}
Furthermore, for $b=3$, we have
\begin{align}\label{C6}
A^{b}(d+b)=q^4-(d+3)q+d+2.
\end{align}
\end{corollary}

{\bf Proof}. Note that
\begin{align*}
\delta_n(n)=1,~~\lfloor M_1(w,b,0)\rfloor=1, (t=0,1)~~\lfloor M_1(w,b,t)\rfloor\le0~(t\ge 2),~~M_2(w,b)=0,
\end{align*}
By Theorem \ref{t31}, we have
\begin{align}\label{S1}
A^{b}(d+b)=
 \sum\limits_{t=0}^{b-1}(t+1)F_{b,d}(d+b-t)+\sum\limits_{\substack{a_1,a_{2}\in \mathbb{N}^{*}\\a_1+a_{2}=d+1}}F_{b,d}(a_1,a_{2})+2\sum\limits_{\substack{a_1,a_{2}\in \mathbb{N}^{*}\\a_1+a_{2}=d}}F_{b,d}(a_1,a_{2}).
\end{align}
Now by Theorem \ref{t32}, one has
\begin{align}\label{S2}
2\sum\limits_{\substack{a_1,a_{2}\in \mathbb{N}^{*}\\a_1+a_{2}=d}}F_{b,d}(a_1,a_{2})=2(d-1)A(d,d).
\end{align}
and
\begin{align}\label{S3}
\begin{aligned}
&\sum\limits_{\substack{a_1,a_{2}\in \mathbb{N}^{*}\\a_1+a_{2}=d+1}}F_{b,d}(a_1,a_{2})\\
=&F_{b,d}(1,d)+F_{b,d}(d,1)+F_{b,d}(2,d-1)+F_{b,d}(d-1,2)+\sum\limits_{\substack{ a_1,a_{2}\in \mathbb{N}^{*}\backslash\{1,2\}\\a_1+a_{2}=d+1}}F_{b,d}(a_1,a_{2})\big)\\
=&2A(d+1,d)+2(d-2)A(d,d)+2A(d+1,d)+2(d-3)A(d,d)\\
&+(d-4)A(d+1,d)\sum\limits_{\substack{ a_1,a_{2}\in \mathbb{N}^{*}\backslash\{1,2\}\\a_1+a_{2}=d+1}}\bigg(\binom{a_1-2}{1}+\binom{a_2-2}{1}\bigg)A(d,d)\\
=&dA(d+1,d)+\big(2(d-2)+2(d-3)+(d-3)(d-4)\big)A(d,d)\\
=&dA(d+1,d)+(d-1)(d-2)A(d,d)
\end{aligned}
\end{align}
Now by $(\ref{S1})$-$(\ref{S3})$, we obtain $(\ref{C5})$.
Furthermore, for $b=3$, we have
\begin{align*}
&F_{3,d}(d+1)=A(d+1,d)+(d-1)A(d,d),\\
&F_{3,d}(d+2)=A(d+2,d)+dA(d+1,d)+\frac{(d-1)(d-2)}{2}A(d,d),\\
&F_{3,d}(d+3)=A(d+3,d)+(d+1)A(d+2,d)+\frac{d(d-1)}{2}A(d+1,d)+\frac{(d-1)(d-2)(d-3)}{6}A(d,d),
\end{align*}
and then 
\begin{align}\label{S4}\begin{aligned}
&\sum\limits_{t=0}^{b-1}(t+1)F_{b,d}(d+3-t)\\
=&F_{3,d}(d+3)+2F_{3,d}(d+2)+3F_{3,d}(d+1)\\
=&A(d+3,d)+(d+3)A(d+2,d)+\frac{d^2+3d+6}{2}A(d+1,d)+\frac{(d-1)(d^2+d+12)}{6}A(d,d).
\end{aligned}
\end{align}
By calculating directly,
\begin{align}\label{S5}
\begin{aligned}
&A(d,d)=q-1,\\
&A(d+1,d)=q^2-(d+1)q+d,\\
&A(d+2,d)=q^3-(d+2)q^2+\frac{(d+2)(d+1)}{2}q-\frac{d(d+1)}{2},\\
&A(d+3,d)=q^4-(d+3)q^3+\frac{(d+3)(d+2)}{2}q^2-\frac{(d+3)(d+2)(d+1)}{6}q+
\frac{d(d+1)(d+2)}{6}.
\end{aligned}
\end{align} 
Now, by $(\ref{C5})$ and $(\ref{S4})$-$(\ref{S5})$, we have
\begin{align*}
&A^{3}(d+3)\\
=&A(d+3,d)+(d+3)A(d+2,d)+\frac{(d+2)(d+3)}{2}A(d+1,d)+\frac{(d-1)(d^2+7d+12)}{6}A(d,d)\\
=&q^4-(d+3)q+d+2.
\end{align*}
$\hfill\Box$

\begin{remark}\label{R2}
For $b=3$ and $d\ge 3$, let $\mathcal{C}$ be an $[d+3,4,3]_q$ MDS code, we know that $A^{3}(0)=1$ and $A^3(w)=0$ $(i=1,\ldots,d+1)$. Furthermore,
By Corollary $\ref{C3}$, we have $A^{3}(d+2)=(d+3)(q-1)$. By Corollary $\ref{C4}$, we have $A^{3}(d+3)=q^4-(d+3)q+d+2.$ Obviously, $A^3(0),\ldots,A^3({d+3})$ are
accordant with $\sum\limits_{w=0}^{d+3}A^3(w)=q^{4}.$
\end{remark}

\begin{example}
	By using Matlab program, we can obtain the $3$-weight distribution for Reed-Solomon code $[6,4,3]_{11}$ as follows
\begin{align*}
A^3(0)=1,~A^3(1)=A^3(2)=A^3(3)=A^3(4)=0,~A^3(5)=60,~A^3(6)=14580,
\end{align*}which is accordant with Remark \ref{R2}.
\end{example}
\section{The proof for Theorem \ref{t31}}
In order to make the proof easier to understand ,we give the necessary notation, and then present the framework of the proof for Theorem $\ref{t31}$ as follows.\\ 

{\bf Notation:}  For a vector $\mathbf{c}\in\mathbb{F}_q^n\backslash\{\mathbf{0}\}$, we decompose the $n$ coordinates of $\mathbf{c}$ into $l+2$ blocks, let
\begin{align*}
[1,n]=N_0\cup N_1\cup N_2\cup\cdots\cup N_l\cup N_{l+1}~~(l\ge 0),
\end{align*}where{\small
\begin{align*}
& N_0=[1,n_0],\\
& N_i=[n_0+\cdots+n_{i-1}+1,n_0+\cdots+n_i]~ (1\le i\le l),\\
& N_{l+1}=[n_1+\cdots+n_l+1,n_1+\cdots+n_l+n_{l+1}],
\end{align*}}with
$|N_0|=n_0\ge 0$, $|N_{l+1}|\ge0$ and $|N_{l}|=n_i\ge 1$ $(1\le i\le l)$. It is easy to see $n=\sum\limits_{i=0}^{l+1}n_i$. Furthermore, for completeness, if $n_0=0$ $(n_{l+1}=0)$, we set  $N_0=\emptyset$ $(N_{l+1}=\emptyset)$.
Now we call a vector $\mathbf{c}=(c_1,\ldots,c_{n})\in\mathbb{F}_q^{n}\backslash\{ \mathbf{0}\}$ with $b$-shape $(N_0,N_1,\ldots,N_{l+1})$ if it satisfies the following conditions.	

$(\mathrm{C}1)$ $n_0=\min\{i~|~c_i\neq0,~ 1\le i\le n\}-1$;

$(\mathrm{C}2)$ $n_{l+1}=n-\max\{i~|~c_i\neq 0,~ 1\le i\le n\}$;

$(\mathrm{C}3)$ For any odd $i$ $(1\le i\le l)$, $c_{n_0+\cdots+n_{i-1}+1}$ and $c_{n_0+\cdots+n_i}$ are both nonzero, and 
\begin{align*}
wt_{b-1}\big((c_{n_0+\cdots+n_{i-1}+1},c_{n_0+\cdots+n_{i-1}+2},\ldots,c_{n_0+\cdots+n_i})\big)=n_i;
\end{align*}

$(\mathrm{C}4)$ For any even $i$ $(1\le i\le l)$, $n_i\ge b-1$ and $(c_{n_0+\cdots+n_{i-1}+1},\ldots,c_{n_0+\cdots+n_i})=\mathbf{0}$.\\

 In the following statement,  we alway assume  $\mathcal{C}$ is an $[n,n+1-d,d]$ MDS code and $|N_i|=n_i$ $(i=0,\ldots,l+1)$. Now we give the framework of the proof for Theorem \ref{t31}.
 
$\bullet$ For any vector $\mathbf{c}\neq 0$ with  $b$-shape $(N_0,N_1,\ldots,N_{l+1})$, it is ovbious that $l$ must be odd. Let $t=n_0+n_{l+1}$, by $(\mathrm{C}1)$-$(\mathrm{C}4)$, we can get
\begin{align}\label{www1}\begin{aligned}
	 wt_b(\mathbf{c})=
\begin{cases}
n,\quad& t\le b-2\text{~and~} l=1;\\
t+\sum\limits_{i=1}^{\frac{l+1}{2}}n_{2i-1}+\frac{(l-1)(b-1)}{2},\quad& t\le b-2\text{~and~} l\ge 3;\\
n-(t+1-b),\quad& t\ge b-1\text{~and~} l=1;\\
\sum\limits_{i=1}^{\frac{l+1}{2}}n_{2i-1}+\frac{(l+1)(b-1)}{2},\quad& t\ge b-1\text{~and~} l\ge 3.
\end{cases}
\end{aligned}
\end{align}

$\bullet$ Basing on (\ref{www1}), we know that the $b$-weight of a vector can be uniquely determined by the $b$-shape. Thus for $t \in \mathbb{N}$, $w,l\in\mathbb{N}^{*}$, the set 
\begin{align*}
\mathrm{B}(w,t,l)=\big\{(N_0,\ldots,N_{l+1})\!~\big|\!~|N_0|+|N_{l+1}|=t,~\text{if}~\mathbf{c}\text{~has~$b$-shape $(N_0,\ldots,N_{l+1})$,~then $wt_b(\mathbf{c})=w$}\big\}
\end{align*}
is well defined.  Let \begin{align*}
 \tilde{A}\big(N_0,\ldots,N_{l+1}\big)=\Big\{\mathbf{c}\in\mathcal{C}~|~\text{$\mathbf{c}$~has $b$-shape $(N_0,\ldots,N_{l+1})$~and~$wt_b(\mathbf{c})=w$}\Big\}.
 \end{align*}
 Then the number of codewords in $\mathcal{C}$ with $b$-weight $w$ is 
 \begin{align*}
 	A^{b}(w)=\sum\limits_{t}\sum\limits_{l}\sum\limits_{(N_0,\ldots,N_{l+1})\in\mathrm{B}(w,t,l)}\tilde{A}\big(N_0,\ldots,N_{l+1}\big).
 \end{align*}
 
 $\bullet$ By shortening $\mathcal{C}$ on $N_0\cup N_2\cup\cdots\cup N_{l+1}$, we prove that $\tilde{A}(N_0,\ldots,N_{l+1})=F_{b,d}(n_1,n_3,\ldots,n_{l})$.
 
 $\bullet$  Basing on $(\ref{www1})$, the calculation of   $\sum\limits_{(N_0,\ldots,N_{2})\in\mathrm{B}(w,t,1)}F_{b,d}(n_1,n_3,\ldots,n_{l})$ is  divided into four cases, and then it converts to calculate the number of solutions for the corresponding equations.\\

Before giving the proof of Theorem $\ref{t31}$, we prove  $(\ref{www1})$ at first.

\begin{lemma}\label{ll0}
	 For any $\mathbf{c}\in\mathbb{F}_q^n\backslash\{\mathbf{0}\}$ with $b$-shape $(N_0,N_1,\ldots,N_{l+1})$,  let $n_i=|N_i|$ and $t=n_0+n_{l+1}$, then
	 \begin{align}\label{wwwc}\begin{aligned}
	 wt_b(\mathbf{c})=
	 \begin{cases}
	 n,\quad& t\le b-2\text{~and~} l=1;\\
	 t+\sum\limits_{i=1}^{\frac{l+1}{2}}n_{2i-1}+\frac{(l-1)(b-1)}{2},\quad& t\le b-2\text{~and~} l\ge 3;\\
	 n-(t+1-b),\quad& t\ge b-1\text{~and~} l=1;\\
	 \sum\limits_{i=1}^{\frac{l+1}{2}}n_{2i-1}+\frac{(l+1)(b-1)}{2},\quad& t\ge b-1\text{~and~} l\ge 3.
	 \end{cases}
	 \end{aligned}
	 \end{align}
\end{lemma}

{\bf Proof}.  For any  $\mathbf{c}\in\mathbb{F}_q^n\backslash\{\mathbf{0}\}$ with  $b$-shape $(N_0,N_1,\ldots,N_{l+1})$, obviously, $l$ is odd.  Now  $wt_{b}(\mathbf{c})$ is derived by $t$ and $n_i$ $(i=1,\ldots,l)$ in the following four cases.

If $t\le b-2$ and $l=1$, then $\mathbf{c}$ has  $b$-shape $(N_0,N_1,N_{2})$ with $|N_0|+|N_2|=t\le b-2$, by condition $(\mathrm{C}{3})$, we have $$wt_{b}(\mathbf{c})=n.$$

If $t\le b-2$ and $l\ge3$, by conditions $(\mathrm{C}3)$-$(\mathrm{C}4)$ and $\sum\limits_{i=0}^{l+1}n_i=n$, one has
\begin{align*}
\begin{aligned}
wt_{b}(\mathbf{c})=&n-\big(n_2-(b-1)\big)-\cdots-\big(n_{l-1}-(b-1)\big)\\
=&n-\sum\limits_{i=1}^{\frac{l-1}{2}}n_{2i}+{\frac{(l-1) (b-1)}{2}}\\
=&t+\sum\limits_{i=1}^{\frac{l+1}{2}}n_{2i-1}+{\frac{(l-1) (b-1)}{2}}.
\end{aligned}	
\end{align*}

If $t\ge b-1$ and $l=1$, then $\mathbf{c}$ has  $b$-shape $(N_0,N_1,N_{2})$ with $|N_0|+|N_2|=t\ge b-1$, now by condition $(\mathrm{C}3)$, we get
\begin{align*}
wt_{b}(\mathbf{c})=n-(t+1-b).
\end{align*}

If $t\ge b-1$ and $l\ge 3$, by conditions $(\mathrm{C}3)$-$(\mathrm{C}4)$ and $\sum\limits_{i=0}^{l+1}n_i=n$, one has
\begin{align*}
\begin{aligned}
wt_{b}(\mathbf{c})=&n-\big(t-(b-1)\big)-\big(n_2-(b-1)\big)-\cdots-\big(n_{l-1}-(b-1)\big)\\
=&n-t-\sum\limits_{i=1}^{\frac{l-1}{2}}n_{2i}+{\frac{(l+1) (b-1)}{2}}\\
=&\sum\limits_{i=1}^{\frac{l+1}{2}}n_{2i-1}+{\frac{(l+1) (b-1)}{2}}.
\end{aligned}	
\end{align*}

$\hfill\Box$


{\bf The proof of Theorem \ref{t31}}. We prove  Theorem \ref{t31} by the following two Steps.

${\bf Step}$ 1. If $n_1+\cdots+n_{l}<d$, it is obviously that $\tilde{A}(N_0,\ldots,N_{l+1})=0$; otherwise, we show that \begin{align}\label{AEF}
	\tilde{A}(N_0,\ldots,N_{l+1})=F_{b,d}(n_1,n_3,\ldots,n_{l}).
\end{align}
Let $\mathcal{C}_{N_0\cup N_2\cup\cdots\cup N_{l+1}}$ be an shorten code of $\mathcal{C}$ on $N_0\cup N_2\cup\cdots\cup N_{l+1}$,  by Lemma \ref{MDSS}, $\mathcal{C}_{N_0\cup N_2\cup\cdots\cup N_{l+1}}$ is an $\Big[~n_1+n_3+\cdots+n_l,~n_1+n_3+\cdots+n_l-d+1,~d~\Big]$ MDS code. Thus $\tilde{A}\big(N_0,\ldots,N_{l+1}\big)$ is equal to the number of  $\mathbf{c}\in\mathcal{C}_{N_0\cup N_2\cup\cdots\cup N_{l+1}}$, which satisfies

$\bullet$ $ \mathbf{c} =(\underbrace{c_1,\ldots,c_{n_1}}_{1},\underbrace{c_{n_1+1},\ldots,c_{n_1+n_3}}_{2},\ldots,\underbrace{c_{n_1+n_3+\cdots+n_{l-2}+1},\ldots,c_{n_1+n_3+\cdots+n_{l-2}+n_{l}}}_{\frac{l+1}{2}})$ ~with\\
$~~~~~~~~~~~~~~~~~~~c_{n_1+n_3+\cdots+n_{2i-1}+1}\neq0~~~~~~\text{and}~~~~~~c_{n_1+n_3+\cdots+n_{2i-1}+n_{2i+1}}\neq 0 \quad\big(i=0,1,\ldots, \frac{l-1}{2}\big).$\\

$\bullet$ $wt_{b-1}(\mathbf{c})=n_1+n_3+\cdots+n_l$.\\
Now by conditions $(\mathrm{F}1)$-$(\mathrm{F}2)$,  $(\ref{AEF})$ holds.\\

${\bf Step}$ 2. We give the value of $\sum\limits_{(N_0,\ldots,N_{l+1})\in\mathrm{B}(w,t,1)}F_{b,d}(n_1,n_3,\ldots,n_{l})$ by the following $4$ cases.

{\bf Case $1$}. For $t\le b-2$ and $l=1$, by  the definition of $B(w,t,1)$, $(\mathrm{C}1)$-$(\mathrm{C}3)$ and Lemma $\ref{ll0}$, we know that $(N_0,\ldots,N_{2})\in B(w,t,1)$ if and only if 
\begin{align*}
	\begin{cases}
&\!\!\!	n_0+n_{2}=t;\\
&\!\!\!		n_1=n-t;\\
&\!\!\!	n_1\ge d, n_{2}\ge 0;\\
&\!\!\!		w=n.
	\end{cases}
\end{align*}
Thus 
\begin{align}\label{key1}\begin{aligned}
\sum\limits_{(N_0,N_1,N_{2})\in\mathrm{B}(w,t,1)}F_{b,d}(n_1)=
	\delta_n(w)(t+1)F_{b,d}(n-t),\quad& \text{~if~}t\le n-d,
	\end{aligned}
\end{align} where
\begin{align*}
\delta_n(w)=\begin{cases}
1,\quad &w=n;\\
0,\quad &\text{otherwise}.
\end{cases}
\end{align*}

{\bf Case $2$}. For $t\le b-2$ and $l\ge3$, by  the definition of $B(w,t,l)$, $(\mathrm{C}1)$-$(\mathrm{C}4)$ and Lemma $\ref{ll0}$, we know that $(N_0,\ldots,N_{l+1})\in B(w,t,1)$ if and only if 
\begin{align}\label{l3}
\begin{aligned}
\begin{cases}
n_0+n_{l+1}=t,\\
n_0+n_1+n_2+\cdots+n_l+n_{l+1}=n,\\
n_1+n_3+\cdots+n_l=w-t-{\frac{(l-1) (b-1)}{2}},\\
n_1+n_3+\cdots+n_l\ge d,\\
n_0\ge0,~n_{l+1}\ge 0,~n_i\ge 1~(i=1,3,\ldots,l),~n_j\ge b-1~(j=2,4,\ldots, l-1),
\end{cases}
\end{aligned}
\end{align}
which is equavient to 
\begin{align}\label{l4}
\begin{aligned}
\begin{cases}
n_0+n_{l+1}=t,\\
(n_2-b+2)+(n_4-b+2)+\cdots+(n_{l-1}-b+2)=n-w+\frac{l-1}{2},\\
n_1+n_3+\cdots+n_l=w-t-\frac{(l-1)(b-1)}{2},\\
n_1+n_3+\cdots+n_l\ge d,\\
n_0\ge0,~n_{l+1}\ge 0,~n_i\ge 1~(i=1,3,\ldots,l),~n_j-b+2\ge 1~(j=2,4,\ldots, l-1).
\end{cases}
\end{aligned}
\end{align} 
Thus 	
\begin{align*}
\begin{aligned}
\begin{cases}
w-t-\frac{(l-1)(b-1)}{2}\ge \frac{l-1}{2}+1,\\
w-t-\frac{(l-1)(b-1)}{2}\ge d,
\end{cases}
\end{aligned}
\end{align*}
and then the upper bound of $\frac{l-1}{2}$ is given as
\begin{align}\label{l6}
\frac{l-1}{2}\le \min \bigg\{ \frac{w-t-1}{b},~\frac{w-t-d}{b-1}\bigg\}=M_1(w,b,t). 
\end{align} 
Now by (\ref{l4})-(\ref{l6}) and Lemma \ref{nc}, we have
\begin{align}\label{key2}
\begin{aligned}
&\sum\limits_{(N_0,\ldots,N_{l+1})\in\mathrm{B}(w,t,l)}F_{b,d}(n_1,n_3,\ldots,n_{l})\\=&
\begin{cases}
(t+1)\binom{n-w+\frac{l-1}{2}-1}{\frac{l-1}{2}-1}\!\!\!\!\!\!\!\!\sum\limits_{ \substack{n_1,n_3,\ldots,n_{l}\in \mathbb{N}^{*}\\n_1+n_3+\cdots+n_{l}=w-t-\frac{(l-1)(b-1)}{2}}}\!\!\!\!\!\!\!\!F_{b,d}(n_1,n_3,\ldots,n_{l}),&\text{~if~}\frac{l-1}{2}\le M_1(w,b,t);\\
0,&\text{~if~}\frac{l-1}{2}> M_1(w,b,t).
\end{cases}
\end{aligned}
\end{align}

{\bf Case} $3$.  For $t\ge b-1$ and $l=1$,  by  the definition of $B(w,t,l)$, $(\mathrm{C}1)$-$(\mathrm{C}4)$ and Lemma $\ref{ll0}$,  $(N_0,\ldots,N_{2})\in B(w,t,1)$ if and only if
\begin{align*}
\begin{cases}
	&\!\!\!	n_0+n_{2}=t;\\
	&\!\!\!		n_1=n-t;\\
	&\!\!\!		w=n-(t+1-b);\\
	&\!\!\!	n_1\ge d,	n_0\ge 0,n_{2}\ge 0.
\end{cases}
\end{align*}
Thus
\begin{align*}
t=n-w+b-1,
\end{align*}
and then  {\small
\begin{align}\label{key3}
\sum\limits_{(N_0,N_1,N_{2})\in\mathrm{B}(w,t,1)}F_{b,d}(n_1)=\begin{cases}
(n-w+b)F_{b,d}(w+1-b),&\text{if~}t=n-w+b-1;\\
0,&\text{if~}t\neq n-w+b-1.
\end{cases}
\end{align}}

{	\bf Case} $4$.  If $t\ge b-1$ and $l\ge 3$, by  the definition of $B(w,t,l)$, $(\mathrm{C}1)$-$(\mathrm{C}4)$ and Lemma $\ref{ll0}$, we know that $(N_0,\ldots,N_{l+1})\in B(w,t,1)$ if and only if 
\begin{align*}
\begin{aligned}
\begin{cases}
n_0+n_{l+1}=t,\\
n_0+n_1+n_2+\cdots+n_l+n_{l+1}=n,\\
n_1+n_3+\cdots+n_l=w-{\frac{(l+1) (b-1)}{2}},\\
n_1+n_3+\cdots+n_l\ge d,\\
n_0\ge0,~n_{l+1}\ge 0,~n_i\ge 1~(i=1,3,\ldots,l),~n_j\ge b-1~(j=2,4,\ldots, l-1),
\end{cases}
\end{aligned}
\end{align*}
which is equavient to 
\begin{align}\label{l24}
\begin{aligned}
\begin{cases}
n_0+n_{l+1}=t,\\
(n_2-b+2)+(n_4-b+2)\cdots+(n_{l-1}-b+2)=n-w-t+b-1+\frac{l-1}{2},\\
n_1+n_3+\cdots+n_l=w-\frac{(l+1)(b-1)}{2},\\
n_1+n_3+\cdots+n_l\ge d,\\
n_0\ge0,~n_{l+1}\ge 0,~n_i\ge 1~(i=1,3,\ldots,l),~n_j-b+2\ge 1~(j=2,4,\ldots, l-1),
\end{cases}
\end{aligned}
\end{align}
thus we have
\begin{align}\label{l25}
\begin{aligned}
\begin{cases}
n-w-t+b-1\ge 0,\\
w-\frac{(l+1)(b-1)}{2}\ge \frac{l-1}{2}+1,\\
w-\frac{(l+1)(b-1)}{2}\ge d.
\end{cases}
\end{aligned}
\end{align}
By (\ref{l25}), the upper bounds for $t$ and $\frac{l-1}{2}$ are given as
\begin{align}\label{ut}
\begin{aligned}
t\le n-w+b-1,	
\end{aligned}
\end{align} 
and
\begin{align}\label{ul}
\begin{aligned}
\frac{l-1}{2}\le \min \bigg\{\frac{w-b}{b},~\frac{w-d}{b-1}-1\bigg\}=M_2(w,b).
\end{aligned}
\end{align} 
Now by  (\ref{l24}) and $(\ref{ut})$-$(\ref{ul})$, we have
{\small \begin{align}\label{key4}
\begin{aligned}
&\sum\limits_{(N_0,\ldots,N_{l+1})\in\mathrm{B}(w,t,l)}F_{b,d}(n_1,n_3,\ldots,n_{l+1})\\=&
\begin{cases}
(t+1)\binom{n-w-t+b+\frac{l-1}{2}-2}{\frac{l-1}{2}-1}\!\!\!\!\!\!\!\!\sum\limits_{ \substack{n_1,n_3,\ldots,n_{l}\in \mathbb{N}^{*}\\n_1+\cdots+n_{l}=w-\frac{(l+1)(b-1)}{2}}}\!\!&\!\!\!\!\!\!F_{b,d}(n_1,n_3,\ldots,n_{l}),\\
&\text{~if~}t\le n-w+b-1~\text{and}~\frac{l-1}{2}\le M_2(w,b);\\
0,&\text{~if~}t>n-w+b-1,~\text{or}~\frac{l-1}{2}> M_2(w,b).\\
\end{cases}
\end{aligned}
\end{align}\small}

So far, by $(\ref{AEF})$-$(\ref{key1})$, $(\ref{key2})$-$(\ref{key3})$ and $(\ref{key4})$, we have
{\small \begin{align*}
A^b(w)=&\delta_n(w) \sum\limits_{t=0}^{b-2}(t+1)F_{b,d}(n-t)\\
&+	\sum\limits_{t=0}^{b-2}(t+1)\sum\limits_{\frac{l-1}{2}=1}^{\lfloor M_1(w,b,t)\rfloor}\binom{n-w+\frac{l-1}{2}-1}{\frac{l-1}{2}-1}\sum\limits_{\substack{n_1,n_3,\ldots,n_{l}\in \mathbb{N}^{*}\\n_1+n_3+\cdots+n_{l}=w-t-\frac{(l-1)(b-1)}{2}}}F_{b,d}(n_1,n_3,\ldots,n_{l})\\
&+(n-w+b)F_{b,d}(w-b+1)\\
&+\sum\limits_{t=b-1}^{n-w+b-1}(t+1)\sum\limits_{\frac{l-1}{2}=1}^{\lfloor M_2(w,b)\rfloor}\binom{n-w-t+b+\frac{l-1}{2}-2}{\frac{l-1}{2}-1}\!\!\!\!\!\!\!\!\!\!\!\!\sum\limits_{ \substack{n_1,n_3,\ldots,n_{l}\in \mathbb{N}^{*}\\n_1+n_3+\cdots+n_{l}=w-\frac{(l+1)(b-1)}{2}}}\!\!\!\!\!\!\!\!\!\!\!\!F_{b,d}(n_1,n_3,\ldots,n_{l})\\
=&\delta_n(w) \sum\limits_{t=0}^{b-2}(t+1)F_{b,d}(n-t)+(n-w+b)F_{b,d}(w-b+1)\\
&+	\sum\limits_{t=0}^{b-2}(t+1)\sum\limits_{i=1}^{\lfloor M_1(w,b,t)\rfloor}\binom{n-w+i-1}{i-1}\sum\limits_{\substack{a_1,a_2,\ldots,a_{i+1}\in \mathbb{N}^{*}\\a_1+a_2+\cdots+a_{i+1}=w-t-i(b-1)}}F_{b,d}(a_1,a_2,\ldots,a_{i+1})\\
&+\sum\limits_{t=b-1}^{n-w+b-1}(t+1)\sum\limits_{i=1}^{\lfloor M_2(w,b)\rfloor}\binom{n-w-t+i+b-2}{i-1}\!\!\!\!\!\!\!\!\!\!\!\!\sum\limits_{ \substack{a_1,a_2,\ldots,a_{i+1}\in \mathbb{N}^{*}\\a_1+a_2+\cdots+a_{i+1}=w-(i+1)(b-1)}}\!\!\!\!\!\!\!\!\!\!\!\!F_{b,d}(a_1,a_2,\ldots,a_{i+1}).
\end{align*}}$\hfill\Box$\\

{\bf The proof for Theorem \ref{t32}.} The key of the proof is to find the connection between $F_{b,d}(L_1,\ldots,L_{I})$ and $N_{b}(r,L)$. The following notations is necessary to calculate $F_{b,d}(L_1,\ldots,L_{I})$. 

${\bf Notations.}$

$\bullet$ For any $({c}_{L_1+\cdots+L_{i-1}+1},\ldots,{c}_{L_1+\cdots+L_{i}})\in\mathbb{F}_q^{L_i}\backslash\{\mathbf{0}\}$,  we decompose the $L_i$ coordinates of $\mathbf{c}$ into $j_i$ blocks, i,e,
\begin{align*}
[L_1+\cdots+L_{i-1}+1,L_1+\cdots+L_i]=N_{i,1}\cup N_{i,2}\cup \cdots \cup N_{i,j_i},
\end{align*} 
where 
\begin{align*}
N_{i,y}=[L_1+\cdots+L_{i-1}+n_{i,1}+\cdots+n_{i,y-1}+1,~L_1+\cdots+L_{i-1}+n_{i,1}+\cdots+n_{i,y}]~~ (y=1,\ldots,j_i),
\end{align*}
and  $|N_{i,y}|=n_{i,y}\ge 1$. Obviously, $L_i=\sum\limits_{y=1}^{j_i}n_{i,y}$. we call  $({c}_{L_1+\cdots+L_{i-1}+1},\ldots,{c}_{L_1+\cdots+L_{i}})\in\mathbb{F}_q^{L_i}\backslash\{\mathbf{0}\}$  possessing  shape $(N_{i,1},N_{i,2},\ldots,N_{i,j_i})$ if it satisfies the following two conditions.

$(\mathrm{B}1)$ For every block $N_{i,y}$, either each coordinate in $N_{i,y}$ is zero or nonzero;

$(\mathrm{B}2)$ For any two adjacent blocks $N_{i,y}$ and $N_{i,y+1}$ , either each coordinate in
$N_{i,y}$ is zero and each coordinate in $N_{i,y+1}$ is nonzero, or vice versa.\\	

$\bullet$ Let $\mathbb{V}_i=\Big\{(c_1,\ldots,c_{L_i})\in\mathbb{F}_q^{L_i}~\big|~c_{1}\neq 0,~c_{L_i}\neq 0\Big\}.$  Note that for a vector $(c_1,\ldots,c_{L_i})\in \mathbb{V}_i$ has shape $(N_{i,1},N_{i,2},\ldots,N_{i,j_i})$, then $j_i$ is odd, thus the following set
\begin{align*}
	P_i\big(j_i,m_i\big)=\Big\{(N_{i,1},N_{i,2},\ldots,N_{i,j_i})~|&~|N_{i,1}|+|N_{i,3}|+\cdots+|N_{i,j_i}|=m_i~\text{and}~\\
	&\text{If~$\mathbf{c}\in\mathbb{V}_i$~has~shape~$(N_{i,1},N_{i,2},\ldots,N_{i,j_i})$}, \text{then~$wt_{b-1}(\mathbf{c})=L_i$}\Big\}.
\end{align*}
is well-defined.

$\bullet$ Let $Q(m_1,\ldots,m_{I},N_{1,1},\ldots,N_{1,j_1},\ldots,N_{I,1},\ldots,N_{I,j_I})$ be the number of codewords
$\mathbf{c}=(\underbrace{c_1,\ldots,c_{L_1}}_{1},\underbrace{c_{L_1+1},\ldots,c_{L_1+L_2}}_{2},\ldots,\underbrace{c_{L_1+\cdots+L_{I-1}+1},\ldots,c_{L_1+\cdots+L_I}}_{I})\in\mathcal{C}$
satisfies $(\mathrm{F}1)$-$(\mathrm{F}2)$ and every
$\big(c_{L_1+\cdots+L_{i-1}+1},\ldots,c_{L_1+\cdots+L_i}\big)$ has the Hamming weigh $m_i$ and shape $\big(N_{i,1},\ldots,N_{1,j_i}\big)$. \\

Now by above notations and conditions $(\mathrm{F}1)$-$(\mathrm{F}2)$, we have
\begin{align}\label{FFF}
\begin{aligned}
&F_{b,d}(L_1,\ldots,L_I)\\
=&\prod_{i=1}^{I}\sum\limits_{j_i}\sum\limits_{m_i}\sum\limits_{(N_{i,1},N_{i,2},\ldots,N_{i,j_i})\in P_i(j_i,m_i)} Q(m_1,\ldots,m_{I},N_{1,1},\ldots,N_{1,j_1},\ldots,N_{I,1},\ldots,N_{I,j_I}).
\end{aligned}
\end{align}

In the following, $Q(m_1,\ldots,m_{I},N_{1,1},\ldots,N_{1,j_1},\ldots,N_{I,1},\ldots,N_{I,j_I})$ is given by shorting $\mathcal{C}$. Let 
\begin{align*}
T=\underbrace{N_{1,2}\cup\cdots \cup N_{1,j_1-1}}_{1}\cup\underbrace{N_{2,2}\cup\cdots \cup N_{2,j_2-1}}_{2}\cup\cdots\cup\underbrace{ N_{I,2}\cup\cdots \cup N_{I,j_I-1}}_{I},
\end{align*}
then 
\begin{align*}
\big|[1,n]\backslash T\big|=m_1+m_2+\cdots+m_I.
\end{align*}
If $m_1+m_2+\cdots+m_I<d$, then 
\begin{align}\label{Q1}
	Q(m_1,\ldots,m_{I},N_{1,1},\ldots,N_{1,j_1},\ldots,N_{I,1},\ldots,N_{I,j_I})=0.
\end{align}
Otherwise, let $\mathcal{C}_T$ be the code shortened on $T$ from $\mathcal{C}$. It follows from Lemma \ref{MDSS} that  
$\mathcal{C}_T$ is an MDS code with parameters
\begin{align*}
\bigg[\sum\limits_{i=1}^{I}m_i,~\sum\limits_{i=1}^{I}m_i-d+1,~d~\bigg].
\end{align*}
Ovbiously, $Q(m_1,\ldots,m_{I},N_{1,1},\ldots,N_{1,j_1},\ldots,N_{I,1},\ldots,N_{I,j_I})$ is equal to the number of codewords in $\mathcal{C}_T$ with Hamming weight $\sum\limits_{i=1}^{I}m_i$,
 By Lemma \ref{MDS}, that is  
\begin{align}\label{Q2}
Q(m_1,\ldots,m_{I},N_{1,1},\ldots,N_{1,j_1},\ldots,N_{I,1},\ldots,N_{I,j_I})=\sum\limits_{j=0}^{\sum\limits_{i=1}^{I}m_i-d}(-1)^j\binom{\sum\limits_{i=1}^{I}m_i}{j}
\bigg(q^{\sum\limits_{i=1}^{I}m_i-d+1-j}-1\bigg).
\end{align}

If $b=2$, then $\mathbf{c}\in \mathcal{C}$ which satisfy  conditions $(\mathrm{F}1)$-$(\mathrm{F}2)$ if and only if  all the coordinates of $\mathbf{c}$ are nonzero, i.e., $L_i=m_i$ and $j_i=1$. Thus one has
\begin{align}\label{FA}
F_{b,d}(L_1,\ldots,L_I)=
\sum\limits_{j=0}^{L-d}(-1)^j\binom{L}{j}\Big(q^{L+1-d-j}-1\Big).
\end{align}

If $b=3$, then the calculation of $\sum\limits_{(N_{i,1},N_{i,2},\ldots,N_{i,j_i})\in P_i(j_i,m_i)}~1$ is given by calculating the number of solutions for the corresponding equations.

For $m_i=L_i$, it is easy to see that  $j_{i}=1$ and
\begin{align}\label{F1}
	\sum\limits_{(N_{i,1},N_{i,2},\ldots,N_{i,j_i})\in P_i(j_i,m_i)}1=1.
\end{align}

For $m_i<L_i$, we have $j_i\ge 3$, and then by the definition of $P_i\big(j_i,m_i\big)$, we know that $(N_{i,1},N_{i,2},\ldots,N_{i,j_i})\in P_i\big(j_i,m_i\big)$ if and only if
  \begin{align*}
 \begin{cases}
 n_{i,1}+n_{i,2}+\cdots+n_{i,j_i}=L_i;\\
 n_{i,1}+n_{i,3}+\cdots+n_{i,j_i}=m_i;\\
 n_{i,y}\in \mathbb{N}^{*},~~n_{i,y}\le b-2~(y~ \text{is even});\\
m_i<L_i,~~ j_i\ge 3\text{~is~odd},
 \end{cases}
 \end{align*}
 where $n_{i,y}=|N_{i,y}|$. Above equations is equivalent to
 \begin{align}\label{njzs}
 \begin{cases}
 n_{i,2}+n_{i,4}+\cdots+n_{i,j_i-1}=L_i-m_i;\\
 n_{i,1}+n_{i,3}+\cdots+n_{i,j_i}=m_i;\\
 n_{i,y}\in \mathbb{N}^{*},~~ n_{i,y}\le b-2~(y~\text{is even});\\
m_i<L_i,~~ j_i\ge 3\text{~is~odd}.
 \end{cases}
 \end{align}
 
Now we give the  bound for $m_i$ and $\frac{j_i-1}{2}$. By $(\ref{njzs})$, we have
 \begin{align*}
 \begin{cases}
 \frac{j_i-1}{2}\le n_{i,2}+n_{i,4}+\cdots+n_{i,j_i-1}=L_i-m_i\le \frac{j_i-1}{2}(b-2);\\
 \frac{j_i+1}{2}\le n_{i,1}+n_{i,3}+\cdots+n_{i,j_i}= m_i;\\
 n_{i,y}\in \mathbb{N}^{*},~~\le n_{i,y}\le b-2~(y~\text{is even});\\
 m_i<L_i,~~ j_i\ge 3\text{~is~odd}.
 \end{cases}
 \end{align*}
Thus, 
\begin{align*}\begin{cases}
\frac{j_i-1}{2}\le \min\{ L_i-m_i,~m_i-1\},\\
L_i- m_i\le\frac{j_i-1}{2}(b-2) \le (m_i-1)(b-2)=(b-2)m_i-(b-2),\\
\end{cases}
\end{align*}
it leads

\begin{align}\label{mm3}
\begin{aligned}
\begin{cases}
\big\lceil\frac{L_i-m_i}{b-2}\big\rceil	\le \frac{j_i-1}{2}\le  \min\{ L_i-m_i,~m_i-1\},\\
\big\lceil \frac{L_i+(b-2)}{b-1}\big\rceil\le m_i\le L_i-1.
\end{cases}
\end{aligned}
\end{align}
Now by $(\ref{mm3})$, Lemma \ref{nc} and the definition of $N_{b}(\frac{j_i-1}{2},L_i-m_i)$, we have
 \begin{align}\label{FFP}\begin{aligned}
&\sum\limits_{(N_{i,1},N_{i,2},\ldots,N_{i,j_i})\in P_i(j_i,m_i)}1\\
=&\begin{cases}
\binom{m_i-1}{\frac{j_i-1}{2}}N_{b}(\frac{j_i-1}{2},L_i-m_i),&\text{~if~}(\ref{mm3})~\text{holds};\\
0,&\text{otherwise}.
\end{cases}
 \end{aligned}
 \end{align}
 So far, by $(\ref{FFF})$-$(\ref{F1})$, $(\ref{FFP})$, we have
 \begin{align*}
F_{b,d}(L_1,\ldots,L_I)
=\begin{cases}
A\Big(\sum\limits_{i=1}^{I}L_i,d\Big),\quad&\text{if}~b=2;\\
 f_b(L_1,m_1)\cdots  f_b(L_I,m_I)A\Big(\sum\limits_{i=1}^{I}m_i,d\Big),\quad&\text{if}~b\ge 3.\\
\end{cases} 
 \end{align*}
 where
 \begin{align*}
 f_b(L_i,m_i)=
\sum\limits_{m_i=\big\lceil \frac{L_i+(b-2)}{b-1}\big\rceil}^{L_I}\sum\limits_{e_i=\big\lceil\frac{L_i-m_i}{b-2}\big\rceil}^{\min\{m_i-1,L_i-m_i\}}\binom{m_i-1}{e_i}N_{b}(e_i,L_i-m_i),
 \end{align*}
 and
 {\begin{align*}
 	A\Big(\sum\limits_{i=1}^{I}m_i,d\Big)=\begin{cases}
 	0, &\sum\limits_{i=1}^{I}m_i<d;\\
 	\sum\limits_{j=0}^{\sum\limits_{i=1}^{I}m_i-d}(-1)^j\binom{\sum\limits_{i=1}^{I}m_i}{j}\bigg(q^{\sum\limits_{i=1}^{I}m_i+1-d-j}-1\bigg), &\text{otherwise}.
 	\end{cases}
 	\end{align*}}
$\hfill\Box$\\

\begin{remark}
	In the proof of Theorem \ref{t32}, For the cases $b=2$ and $b\ge 3$, we can see that the method for giving the value of  $F_{b,d}(L_1,\ldots,L_I)$ is different. In fact,
	for $b=2$, the codeword $\mathbf{c}\in \mathcal{C}$ which satisfy  conditions $(\mathrm{F}1)$-$(\mathrm{F}2)$ is equivalent to  all the coordinates of $\mathbf{c}$ are nonzero, i.e., the Hamming weight of $\mathbf{c}$ is equal to the length, and then  $F_{b,d}(L_1,\ldots,L_I)$ can be obtained from the weight distribution of MDS code explictly. However, for the general case $b\ge 3$, it might have zero in  the coordinates of codeword $\mathbf{c}\in \mathcal{C}$ which satisfy  conditions $(\mathrm{F}1)$-$(\mathrm{F}2)$. In order to give the value of $F_{b,d}(L_1,\ldots,L_I)$, we need to establish the connection between $F_{b,d}(L_1,\ldots,L_I)$ and $N_{b}(e_i,L_i-m_i)$ by using the shorten method, which leads the form of the value $F_{b,d}(L_1,\ldots,L_I)$  is more complicated than that of the case $b=2$.
\end{remark}
\section{Conclusions and Further Study}
In this paper, we obtain the $b$-weight distribution for MDS codes, which is a class
of MDS $b$-symbol codes. For an MDS code $\mathcal{C}$, by calculating the number of solutions for some equations and utilizing its shortened codes, we give the connection between the $b$-weight distribution and the number of codewords in shortened codes of $\mathcal{C}$  with special shape. Furthermore, note thet shortened codes of $\mathcal{C}$ are MDS codes, and then the number of these codewords with special shape are obtained by the shorten method. Our result generalizes Theorem $1$ in \cite{ML}.

The key points of our method is that the Hamming weight distribution for an MDS code are uniquely determined by its parameters, and a shorten code of an MDS code is also an MDS code. However, these points are not always true for an MDS $b$-symbol code, but not an MDS code. Thus for an general MDS $b$-symbol code, we can not determine its $b$-weight distribution by using the method in this paper.

\end{document}